\documentstyle[epsfig]{czjphys}         % for LaTeX 2.09
\begin{document}
\title{Uncertainties in the $0\nu\beta\beta$--decay
nuclear matrix elements}  
%
          % Group together all authors with the same affiliation
          % & address and complete the commands `\author' and
          % `\address' subsequently for i, ii,..., vi.
          % Leave the remaining items untouched.
\authori{Vadim Rodin, Amand Faessler and 
Fedor \v Simkovic\footnote{On  leave of absence from Department of Nuclear 
Physics, Comenius University, Mlynsk\'a dolina F1, SK--842 15 
Bratislava, Slovakia}}      
\addressi{Institute f\"{u}r Theoretische Physik der Universit\"{a}t 
T\"{u}bingen, D-72076 T\"{u}bingen, Germany}
\authorii{Petr Vogel}     \addressii{Kellogg Radiation Laboratory 106-38, California Institute 
of Technology, Pasadena, CA 91125, USA and Physics Department, Stanford University, Stanford, CA 94305, USA}
\authoriii{}    \addressiii{}
\authoriv{}     \addressiv{}
\authorv{}      \addressv{}
\authorvi{}     \addressvi{}
%
%Page headings:
\headauthor{ et al.}            % page heading on the even pages
\headtitle{ \ldots}             % page heading on the odd pages
\lastevenhead{ et al.:  \ldots} % p. h. on the last page if even
 % max. 2 codes of Physics and Astronomy Classification Scheme
 % lowercase letters
%%%%%%%%%%%%%% FOR EDITORIAL USE ONLY!!! %%%%%%%%%%%%%%%
\refnum{A}%\total{}\type{}
\daterec{XXX}    %;\\ final version }
\issuenumber{0}  \year{2001}
\setcounter{page}{1}
%\firstpage{1}
%\lastpage{000}
%\makefirsttitle
%%%%%%%%%%%%%%%%%%%%%%%%%%%%%%%%%%%%%%%%%%%%%%%%%%%%%%%%
\maketitle

\begin{abstract}

The nuclear matrix elements $M^{0\nu}$  of the 
neutrinoless double beta decay ($0\nu\beta\beta$)  
of most nuclei with known $2\nu\beta\beta$-decay rates are systematically evaluated 
using the Quasiparticle Random Phase Approximation 
(QRPA) and Renormalized QRPA (RQRPA).  
The experimental $2\nu\beta\beta$-decay rate is used  
to adjust the most relevant parameter, the strength of  
the particle-particle interaction.  
With such procedure the  
$M^{0\nu}$ values become essentially independent on     
single-particle basis size, the axial vector  
quenching factor, etc. 
Theoretical arguments in favor of the adopted way of determining 
the interaction parameters are presented. 
It is suggested that most of the spread among 
the published $M^{0\nu}$'s
can be ascribed to the choices of implicit and explicit parameters,
inherent to the QRPA method.

\end{abstract}

\section{Introduction} 
 
The observation of $0\nu\beta\beta$ decay would immediately tell us that 
neutrinos are massive Majorana particles 
(for reviews see~\cite{FS98,V02,EV2002,EE04}; 
the issues particularly relevant for the program of $0\nu\beta\beta$ decay 
search are discussed in \cite{nustudy}). But without accurate calculations 
of the nuclear matrix elements quantitative 
conclusions about the absolute neutrino masses and mass hierarchies 
can barely be reached.  
Despite years of effort there is at present a lack of consensus  
among nuclear theorists how to correctly calculate the nuclear matrix 
elements, and how to estimate their uncertainty (see e.g. \cite{EE04,bah04}). 
Since an overwhelming majority of published calculations is based 
on the Quasiparticle Random Phase Approximation (QRPA) and its modifications, 
it is worthwhile to try to see what causes the sizable spread of the 
calculated $M^{0\nu}$ values. Does it reflect some fundamental uncertainty, 
or is it mostly related to different choices of various adjustable 
parameters? If the latter is true (and we believe it is) can one 
find and justify an optimal choice that largely removes such  
unphysical dependence? 
 
In our recent papers \cite{Rod03a,Rod06a} we have shown that by adjusting the 
most important parameter, the strength of the particle-particle 
force so that the known rate of the $2\nu\beta\beta$-decay is correctly 
reproduced, the dependence of the calculated $0\nu\beta\beta$ 
nuclear matrix elements $M^{0\nu}$ on the things which are not {\em a priori}
fixed, in particular, the number of included single particle states, 
the different realistic representations of the nucleon $G$-matrix, the axial vector  
quenching factor etc., is essentially removed. 
The method has systematically been applied to calculate the nuclear matrix elements $M^{0\nu}$ 
for most of the nuclei with known experimental $2\nu\beta\beta$-decay rates and arguments in favor of the chosen  
calculation method have been given.

In this contribution to the MEDEX'05 we briefly review the ideas and the results of~\cite{Rod03a,Rod06a}.

\section{Details of the calculation of $0\nu\beta\beta$ decay matrix elements} 
 
Provided that a virtual light Majorana neutrino with  the effective mass  
$\langle m_{\beta\beta} \rangle$, 
\begin{equation} 
\langle m_{\beta\beta} \rangle = \sum_i^N |U_{ei}|^2 e^{i\alpha_i} m_i ~, 
~({\rm all~} m_i \ge 0)~, 
\end{equation} 
is exchanged between the nucleons 
the half-life of the $0\nu\beta\beta$ decay is given by  
\begin{equation} 
\frac{1}{T_{1/2}} = G^{0\nu}(E_0,Z) |{M'}^{0\nu}|^2  
|\langle m_{\beta\beta} \rangle|^2~, 
\end{equation} 
where $G^{0\nu}(E_0,Z)$ is an accurately calculable phase-space factor, 
and ${M'}^{0\nu}$ is the corresponding nuclear matrix element.  
Thus, obviously, any uncertainty in ${M'}^{0\nu}$ makes the value 
of $\langle m_{\beta\beta} \rangle$ equally uncertain. 
 
The elements of the 
mixing matrix $|U_{ei}|^2$ and the mass-squared differences $\Delta m^2$ 
can be determined in oscillation experiments.  
If the existence of the $0\nu\beta\beta$ decay is proved and the  
value of $T_{1/2}$ is found, combining the knowledge of 
$|U_{ei}|^2$ and $\Delta m^2$, a relatively narrow range of 
absolute neutrino mass scale can be determined, in most situations independently of the 
Majorana phases $\alpha_i$~\cite{EV2002}.
 
The nuclear matrix element ${M'}^{0\nu}$ is defined as 
\begin{equation} 
{M'}^{0\nu} = \left(\frac{g_A}{1.25}\right)^2~\langle f  
| -\frac{M^{0\nu}_F}{g^2_A} + M^{0\nu}_{GT} +  
M^{0\nu}_T |i\rangle  
\label{eq:m0nudef} 
\end{equation} 
where $|i\rangle, ~(|f\rangle )$ are the wave functions of the ground 
states of the initial (final) nuclei. The explicit forms of the operators 
$M^{0\nu}_F, M^{0\nu}_{GT}$ and 
$M^{0\nu}_T$ are given in Ref.~\cite{Rod06a,si99}.
We note that for $g_A=1.25$  
nuclear matrix element ${M'}^{0\nu}$ coincides with ${M}^{0\nu}$ 
of our previous work \cite{Rod03a}. This parameterization is chosen so 
that we could later modify the value of $g_A$ and still 
use the same phase space factor $G^{0\nu}(E_0,Z)$  
that contains $g_A^4 = (1.25)^4$, 
tabulated e.g. in Ref.~\cite{si99}.

In~\cite{Rod03a,Rod06a} the quasiparticle random phase approximation (QRPA) 
and its modification, the renormalized QRPA (RQRPA), are used to describe  
the structure of the intermediate nuclear states virtually excited in the  
double beta decay. We stress  
that in the QRPA and RQRPA one can include essentially unlimited set  
of single-particle (s.p.) states, but only a limited subset of configurations  
(iterations of the particle-hole, 
respectively two-quasiparticle configurations),  
in contrast to the nuclear shell model. 
On the other hand, within the QRPA it is not obvious 
how many single particle states one should include. 
Hence, various authors choose this crucial number basically 
for reasons of convenience.

It is well known that the residual interaction is an effective interaction  
depending on the size of the 
s.p. basis. Hence, when the basis is changed, the interaction 
should be modified as well.  
For each nucleus in question three single-particle bases are chosen  
in Refs.~\cite{Rod03a,Rod06a} 
with the smallest set corresponding to $1 \hbar\omega$ particle-hole excitations,  
and the largest   
to about $4 \hbar\omega$ excitations. The s.p. energies are calculated 
with the Coulomb corrected Woods-Saxon potential. 
The in Ref.~\cite{Rod06a} calculations have been performed using $G$-matrix based only on the Bonn-CD  
nucleon-nucleon potential because as it was shown in Ref.~\cite{Rod03a} a particular choice  
of the realistic residual two-body interaction potential  has almost no impact on  
the finally calculated mean value and variance $\sigma$ 
of ${M'}^{0\nu}$, with the overwhelming contribution to $\sigma$ coming 
from the choice of the single-particle basis size. 

In QRPA and RQRPA there are three important global parameters  
renormalizing the bare residual interaction. 
First, the pairing part of the interaction is multiplied by a factor $g_{pair}$ 
whose magnitude is adjusted, for both protons and neutrons 
separately, such that the pairing gaps for the initial and final nuclei  
are correctly reproduced. 
This is a standard procedure and  
it is well-known that within the BCS method the strength of  
the pairing interaction depends on the size 
of the s.p. basis. 
 
Second, the particle-hole interaction block is renormalized by an overall strength 
parameter $g_{ph}$ which is typically adjusted by requiring 
that the energy of the giant GT resonance is correctly reproduced.  
We find that the calculated energy of the giant GT state is almost independent  
of the size of the s.p. basis and is well reproduced with $g_{ph} \approx 1$.  
Accordingly, we use $g_{ph} = 1$ throughout, without adjustment. 
 
Third, an very important strength parameter $g_{pp}$ renormalizes  
the particle-particle interaction 
(the importance of the particle-particle interaction 
for the $\beta\beta$ decay was recognized first in \cite{VZ86}).  
The decay rate for both modes of $\beta\beta$ decay is well known to depend sensitively 
on the value of $g_{pp}$
(in $J^\pi=1^+$ channel the sensitivity originates as a pronounced effect
of variation of the degree of the SU(4)-symmetry violation by the particle-particle
interaction~\cite{VZ86,Rodin05}). 
This property has been used in~\cite{Rod03a,Rod06a} to fix the value of $g_{pp}$ for each of the s.p. bases 
so that the known half-lives of the $2\nu \beta\beta$ decay are correctly reproduced.  
Such an adjustment of $g_{pp}$, 
when applied to all multipoles $J^{\pi}$, has been shown in~\cite{Rod03a,Rod06a} 
to remove much of the sensitivity to the number of s.p. states,  
to the $NN$ potential employed, and even to whether RQRPA or just simple 
QRPA methods are used. This is in contrast to typical conclusion made in the recent past 
(see, e.g.,~\cite{CS03,EE04}) that the values of ${M'}^{0\nu}$ vary substantially depending  
on all of these things. 
 
We believe that the $2\nu$ decay rate is especially suitable for such an adjustment, 
in particular because it involves the same initial and final states as the 
$0\nu$ decay. Moreover, the QRPA is a method designed to describe collective states as 
well as to obey various sum rules. Both double-beta decay amplitudes, 
$0\nu\beta\beta$ and $2\nu\beta\beta$, receive contributions from many  
intermediate states 
and using one of them for fixing parameters of QRPA seems preferable. 
 
It is well known that the calculated Gamow-Teller strength is larger 
than the experimental one. Formally, this quenching could be conveniently accomplished 
by replacing the true value of the axial current coupling constant $g_A$ = 1.25 by a quenched 
value $g_A \simeq$ 1.0. To see the dependence on 
the chosen $g_A$ value, both values of $g_A$ have been used (for all multipoles). 
The matrix elements ${M'}^{0\nu}$ calculated for the three 
s.p. bases and a fixed $g_A$ are relatively close to each other. 

For each nucleus the corresponding 
average  $\langle {M'}^{0\nu} \rangle$ matrix elements  
(averaged over the three choices of the s.p. space)  
is evaluated, as well as its variance $\sigma$. The errors induced in $\langle {M'}^{0\nu} \rangle$
by the experimental uncertainties in  $M^{2\nu}_{exp}$ are added to the theoretical ones.  
 
\begin{figure}[!t]
\centerline{\epsfig{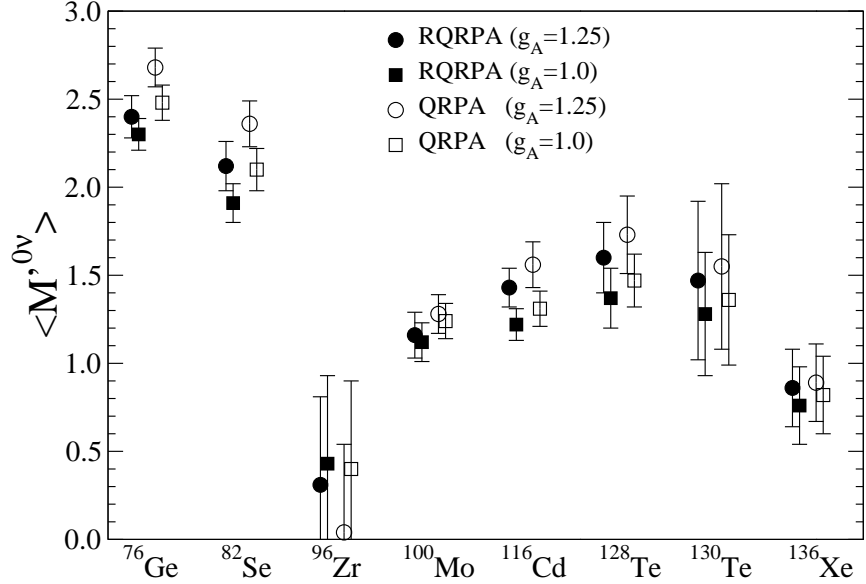}}
\caption{Average nuclear matrix elements $\langle {M'}^{0\nu} \rangle $ 
and their variance (including the error coming from the experimental uncertainty in $M^{2\nu}$) 
for both methods and for all considered nuclei. For $^{136}$Xe the error bars encompass the whole interval related 
to the unknown rate of the $2\nu\beta\beta$ decay.
\label{0nbbfig:2}}
\end{figure}

The calculated $0\nu\beta\beta$ matrix elements are presented in  Fig.~\ref{0nbbfig:2}. There 
the averaged nuclear matrix elements for both methods and both choices of $g_A$ are  
shown along with their full uncertainties (theoretical plus experimental). 
One can see that not only is the variance  
substantially less than the average value, but the results of QRPA  
are quite close to the RQRPA values.  Furthermore, the ratio of the  
matrix elements calculated with different $g_A$ 
is closer to unity (in most cases they differ only by $\sim$20\%)  
than the ratio of the respective $g_A$ squared (1.6 in our case).  
Again, such a partial compensation of the $g_A$-dependence has its origin
in the adopted way of fixing $g_{pp}$ to reproduce $M^{2\nu}_{exp}$
because one needs smaller $g_{pp}$ for smaller $g_A$.

One can qualitatively understand why our chosen procedure stabilizes 
the ${M'}^{0\nu}$ matrix elements as follows: $M^{2\nu}$ matrix elements 
involve only the $1^+$ states in the intermediate odd-odd 
nucleus. The contributions 
of the $1^+$ multipole for both modes of the $\beta\beta$ decay 
depend very sensitively on $g_{pp}$ due to a proximity to 
the point of the phase-transition in the $1^+$ channel (corresponding 
to the collapse of the QRPA equations of motion). 
On the other hand, virtual excitations of many states of different multipolarities 
in the intermediate odd-odd nucleus contribute to
the  ${M'}^{0\nu}$ matrix element, due to the presence of the neutrino 
propagator. 
The multipoles, other than $1^+$, correspond 
to small amplitudes of the collective motion; there is no instability 
for realistic values of $g_{pp} \sim 1.0$. Hence, 
they are much less sensitive to the value of $g_{pp}$. 
By making sure that the contribution of the $1^+$ multipole is fixed, we  
therefore stabilize the 
${M'}^{0\nu}$  value. The fact that RQRPA essentially removes the 
instability becomes then almost irrelevant thanks to the  
chosen adjustment of $g_{pp}$.    
 
One can see in Fig. 2 of \cite{Rod06a} that all multipoles $J^{\pi}$, 
with the exception of the $1^+$ and various 
very small entries, contribute with the same sign and show the essential  
stability of the partial contributions against variation of the basis size. 
This suggests that 
uncertainties in one or few of them will have relatively minor effect. 
It is instructive to see separately the effect of the sometimes
neglected short-range repulsive nucleon-nucleon repulsion and of
the induced weak nucleon currents. One can see in Fig. 6 of~\cite{Rod06a} that the conclusion
of the relative role of different multipoles is affected by these
terms. For example, in $^{100}$Mo the $1^-$ multipole is the strongest 
one when all effects are included, while the $2^-$ becomes dominant
when they are neglected.

\section{Uncertainties of the $0\nu\beta\beta$ decay matrix elements} 

Ideally, the chosen nuclear structure method should describe 
all, or at least very many, experimental data and do that without 
adjustments. That is not the case of the QRPA 
or the RQRPA. The interaction used is an effective interaction, 
and various parameters are adjusted.
One of our goals in~\cite{Rod06a} was to show that a majority of differences among various 
QRPA-like calculations can be understood and possible convergence of the QRPA results can be discussed. 

A detailed list of 13 main reasons  
leading to a spread of the published QRPA and RQRPA results is given in~\cite{Rod06a}. 
Some nuclei, like $^{100}$Mo, exhibit more sensitivity to these  
effects than others, such as $^{76}$Ge or $^{82}$Se. That is confirmed
by our numerical studies.  
The two very important reasons are: \\
1) The choice of $g_{pp}$ (usually fixed to reproduce the experimental half-lives either
of $\beta$-decays or of $2\nu\beta\beta$-decay)\\
2) Whether the two-nucleon short-range correlations (s.r.c.)
are taken into account (for realistic $g_{pp}$ values neglecting s.r.c. would lead to a twofold increase in $M^{0\nu}$).

A moderate spread (which can be as much as tens of \%) can originate from the difference 
choices of the mean field, many-body approximations (RQRPA, SC-QRPA etc.), the size of the model space,
the residual nucleon-nucleon interaction (schematic zero-range or realistic interactions based on the $G$-matrix),
renormalization of the axial-vector coupling constant $g_A=1.0 \div 1.25$ and the
higher order terms (h.o.t.) of the nucleon current (induced pseudoscalar and weak magnetism, about 
$30\%$ reduction in $M^{0\nu}$).

%Still unresolved issues which can significantly affect $M^{0\nu}$
%influence of the nuclear deformation, application of the continuum-QRPA, overlap factor of intermediate nuclear states.

The differences between many calculations are understandable just from the way $g_{pp}$ was fixed, 
the considered size of the model space, the inclusion of the s.r.c. and other minor effects. 

Based on our analysis, we suggest that it is  not appropriate to treat all  
calculated $0\nu\beta\beta$-decay matrix elements at the same level,  
as it is commonly done (see e.g., \cite{bah04,CS03}), and to estimate their 
uncertainty based on their spread. 
Clearly, when some authors do not include effects that 
should be included (e.g. the s.r.c. or the h.o.t. in the nucleon current) 
their results should be either corrected or neglected. 
Some effects are correlated, like the size of the model space 
and the renormalization of the particle-particle interaction. Again, 
if those correlations are not taken into account, erroneous conclusion might 
be drawn. In our works \cite{Rod03a,Rod06a}
we have shown that our way of fixing the model parameters 
removes, or at least greatly reduces, the dependence of the final result 
on most of the effects described above. 

Even after all relevant parameters have been carefully fixed, the QRPA is not able 
always to describe well all relevant 
weak transitions. In particular, it is sometimes impossible to describe 
simultaneously the $2\nu\beta\beta$ decay rate as well as 
the $\beta^-$ and $\beta^+/EC$ matrix elements connecting the 
$1^+$ ground states of the intermediate nucleus  
with the ground states of the final and initial nuclei 
($^{100}$Mo is a well known example of this problem, see e.g.~\cite{grif92}). 
That is an obvious drawback of the 
QRPA method; it is never meant to describe in detail properties 
of non-collective states. But that is less relevant 
for the description of integral quantities that depend on  
sums over many states. 
 
The calculations of the $0\nu\beta\beta$-decay matrix elements  
by Civitarese and Suhonen \cite{CS03} 
deserve more comments. The authors performed them  
within the approach suggested  
in~\cite{suh91} which employs the nucleon  
current derived from the quark wave functions. {\bf 
It is very important to note that in this approach the two  
nucleon s.r.c. are not taken into account.}
The h.o.t. of nucleon current 
were studied~\cite{suh91} with the conclusion of their minor role (in contrast to our one).
 
The extension of the work~\cite{suh91} for the case  
when $g_{pp}$ is adjusted to reproduce the 
single $\beta$-decay amplitudes was presented 
in~\cite{au98}. In~\cite{CS03} the  
nuclear matrix elements are calculated in the same way as in 
\cite{au98}, however, the obtained results differ significantly 
(see Table II of~\cite{Rod06a}) from each other. For some 
nuclei the difference  is as large as a factor of two. There is 
no discussion of this there or in the later Refs.~\cite{CS03,Suh04}. 
It is noteworthy that the largest matrix element in~\cite{CS03} is 
found for the $0\nu\beta\beta$-decay of $^{136}$Xe that 
disagrees with the results of other authors. 
The reduction of the $0\nu\beta\beta$-decay 
of the $^{136}$Xe is explained by the closed neutron shell  
for this nucleus: a sharper Fermi surface leads to a reduction 
of this transition. 
 
Altogether, the matrix elements of \cite{au98,CS03} are noticeably larger
than the present ones. Most of that difference can be attributed to the neglect of the 
s.r.c. and of the h.o.t. of the nucleon weak current in these papers.

In summary of this section we list the main arguments why we believe that
the procedure of adjustment used in our works~\cite{Rod03a,Rod06a} ($g_{pp}$ 
from the $2\nu\beta\beta$-decay) is preferable to
the procedure advocated in \cite{Suh04} ($g_{pp}$ from the single $\beta$-decay) by Suhonen
and in~\cite{oaj} by Suhonen and Civitarese:

\begin{list}{}{\leftmargin=\parindent \rightmargin=0pt 
\itemindent=-\leftmargin \itemsep=-5pt} 
\item i) The QRPA is never meant to describe in detail properties
  of non-collective states like the single beta decay of the
  ground state of a nucleus. Thus, it is more
  relevant to fix the QRPA parameters using an integral quantity like
  the $2\nu\beta\beta$-decay half-life and not the beta decay of a single state.
\item ii) We have showed in \cite{Rod06a} that the first $1^+$ state of the intermediate
  nucleus is not the only one contributing to the $2\nu\beta\beta$-decay amplitude. 
  For instance, in A=76 system  many excited intermediate $1^+$ states give comparable contributions
(Fig.~8 of~\cite{Rod06a}). 
Thus, to give preference to the lowest state is not well-justified, the entire sum is  
actually what matters.
\item iii) The contribution of the $1^+$ multipole to the $M^{0\nu}$ and the corresponding $M^{2\nu}$ 
are correlated (Fig.~8 of~\cite{Rod06a}). Making sure that the $M^{2\nu}$ agrees with its experimental value 
constrains the $1^+$ part of the $M^{0\nu}$ as well. 
The sensitivity of $1^+$ multipole component of the $M^{0\nu}$
to $g_{pp}$ far exceeds sensitivity of the other multipoles. 
Hence, making sure that the whole $1^+$ multipole is correct is crucial.
\item iv) The approach of fixing $g_{pp}$ to reproduce the $\beta^-$ decay of the first
  $1^+$ state of the intermediate nucleus is limited only for
  nuclei with A=100, 116 and 128, where the ground state of
  the corresponding odd-odd nucleus has $1^+$. 
  At the same time, the $2\nu\beta\beta$-decay 
  half-life is known now practically for all nuclei of experimental interest,
  i.e., one can use it to fix $g_{pp}$.
\item v) The running sum contributions 
to the $M^{0\nu}$ in  $^{76}$Ge and $^{100}$Mo (Figs.~8,9 of~\cite{Rod06a}), 
for the different multipoles as well as for the total matrix element
do not reveal a single state dominance. 
Thus, it is not obvious that it is best to choose any 
particular state or transition for the adjustment. 

\item vi) Finally, adjusting $g_{pp}$ to reproduce the $2\nu\beta\beta$-decay 
  half-life essentially removes the dependence on other parameters.
 While we have demonstrated that for all 9 nuclear systems, a similar proof was not given in \cite{Suh04}. 

\end{list}

It clearly follows from the above analysis that the Table 3 of \cite{Suh04} does not reflect
real physical situation and therefore should be disregarded.  

In the very recent publications of Civitarese and Suhonen \cite{oaj} 
serious shortcomings are claimed in the procedure of fixing $g_{pp}$ adopted by us. 
Their criticism is based on the consideration of
negative values of the $2\nu\beta\beta$-decay matrix elements
compatible with the data. But they have overlooked that if 
$g_{pp}$ is fixed to reproduce the single $\beta$-decay data, the problem of 
negative values of the corresponding matrix elements is present as well. 
In~\cite{Rod06a} this criticism of~\cite{oaj} has been 
refuted by physical arguments which explain why 
solutions corresponding to the negative values of $2\nu\beta\beta$-decay 
should not be considered.

It is also worth to mention that in~\cite{oaj}
the calculations have been performed in the QRPA without consideration 
of the effect of two-nucleon short-range correlations. The higher multipolarities
are strongly suppressed by the s.r.c. and h.o.t. in the nucleon current. 
Thus there is no more a clear dominance of the 
$2^-$ multipolarity found in~\cite{oaj}. Of course, the inclusion of
the s.r.c. and the induced pseudoscalar coupling 
of nucleon current in the calculation of the $M^{0\nu}$
is a question of physics and not just a matter of taste. 

In addition, a new factor $(m_e R)^{-2}$ introduced without any justification 
by Civitarese and Suhonen \cite{oaj} into the definition of the $M^{0\nu}$
will definitely contribute to the confusion among the experimentalists working in the field.

\section{Summary and conclusions}

We have shown that the procedure suggested in our recent works, Refs.~\cite{Rod03a,Rod06a}, 
is applicable to essentially all nuclei with known $2\nu\beta\beta$ decay lifetimes.  
Adjusting the strength 
of the particle-particle neutron-proton force $g_{pp}$ in such a way that the 
experimental $2\nu\beta\beta$ decay rate is correctly reproduced removes much 
of the dependence on the size of the single-particle basis and  
whether QRPA or RQRPA is used. We have also shown that the quenching of the axial 
current matrix elements, parameterized by the reduction of the coupling constant 
$g_A$, also leaves the resulting $0\nu\beta\beta$ matrix elements almost unchanged; 
they become insensitive to the variations of parameters describing the short-range 
nucleon-nucleon correlations as well. Thus,  
the resulting $0\nu\beta\beta$ matrix elements acquire well defined values, 
free of essentially arbitrary choices.  
We also present arguments while we believe 
that the chosen procedure of adjusting the interaction is preferable to other 
proposed ways of adjustment. 
 
The differences in most, albeit not all, published QRPA and RQRPA results can be understood. 
Comparison between the results of different QRPA/RQRPA calculations
would be facilitated
if authors of future publications specify in detail what choices of
explicit and implicit adjustable parameters 
they made, and discuss the dependence of their result on their particular 
choice. By following these 
suggestions a consensus among the practitioners of QRPA/RQRPA
could be reached and most of the spread between the calculated nuclear matrix elements, 
that causes much confusion in the wider physics community, would be shown 
to be essentially irrelevant. To reach a convergence
of the results obtained using QRPA/RQRPA is clearly just an important step on the way
to reliable and correct $0\nu\beta\beta$ decay nuclear matrix elements.

The work of F. \v{S}. and V. R. was supported in part by the Deutsche 
Forschungsgemeinschaft (grants 436 SLK 17/298, TU 7/134-1 and FA67/28-2,
respectively). 
We thank also  the EU ILIAS project under the contract RII3-CT-2004-506222. 
The work of P. V. was  supported by the U.S. DOE, Stanford University, SLAC and KIPAC.

\end {document}